\documentclass[aps,prb,twocolumn,amssymb,superscriptaddress]{revtex4-2}

\usepackage{graphicx}
\usepackage{amsmath}
\usepackage{xcolor}
\usepackage[pagewise]{lineno}
\usepackage[colorlinks]{hyperref}
\definecolor{Blue}{rgb}{.0,.17,.5}
\definecolor{Levand}{rgb}{.52,.0,.29}
\hypersetup{
	colorlinks = true,
	citecolor = {Levand},
	linkcolor = {Blue},
	urlcolor = {Blue},
}
\usepackage{fancyhdr}
\begin{document}


\title{Interplay of voltage control of magnetic anisotropy, spin transfer torque, and heat in the spin-orbit torque switching in three-terminal magnetic tunnel junctions}


\author{Viola Krizakova}
\email[]{viola.krizakova@mat.ethz.ch}
\affiliation{Department of Materials, ETH Zurich, 8093 Zurich, Switzerland}

\author{Eva Grimaldi}
\affiliation{Department of Materials, ETH Zurich, 8093 Zurich, Switzerland}

\author{Kevin Garello}
\affiliation{imec, Kapledreef 75, 3001 Leuven, Belgium}
\affiliation{CEA, SPINTEC, F-38000 Grenoble, France}

\author{Giacomo Sala}
\affiliation{Department of Materials, ETH Zurich, 8093 Zurich, Switzerland}

\author{Sebastien Couet}
\affiliation{imec, Kapledreef 75, 3001 Leuven, Belgium}

\author{Gouri Sankar Kar}
\affiliation{imec, Kapledreef 75, 3001 Leuven, Belgium}

\author{Pietro Gambardella}
\email[]{pietro.gambardella@mat.ethz.ch}
\affiliation{Department of Materials, ETH Zurich, 8093 Zurich, Switzerland}


\date{\today}

\begin{abstract}
We use three-terminal magnetic tunnel junctions (MTJs) designed for field-free switching by spin-orbit torques (SOTs) to systematically study the impact of dual voltage pulses on the switching performances. We show that the concurrent action of an SOT pulse and an MTJ bias pulse allows for reducing the critical switching energy below the level typical of spin transfer torque while preserving the ability to switch the MTJ on the sub-ns time scale. By performing dc and real-time electrical measurements, we discriminate and quantify three effects arising from the MTJ bias: the voltage-controlled change of the perpendicular magnetic anisotropy, current-induced heating, and the spin transfer torque. The experimental results are supported by micromagnetic modeling. We observe that, depending on the pulse duration and the MTJ diameter, different effects take a lead in assisting the SOTs in the magnetization reversal process. Finally, we present a compact model that allows for evaluating the impact of each effect due to the MTJ bias on the critical switching parameters. Our results provide input to optimize the switching of three-terminal devices as a function of time, size, and material parameters.
\end{abstract}


\maketitle
\thispagestyle{fancy}
\section{Introduction}

\lfoot{{\footnotesize This article may be downloaded for personal use only. Any other use requires	prior permission of the author and APS Physics Press Office. This article appeared in Physical Review Applied and can be found at: \href{https://journals.aps.org/prapplied/abstract/10.1103/PhysRevApplied.15.054055}{10.1103/PhysRevApplied.15.054055}.}}

The ability to reverse the magnetization direction in a fast and energetically efficient way is a key requirement for magnetic tunnel junctions (MTJs) used in magnetoresistive random-access memories (MRAMs). Current-induced spin transfer torques (STT) \cite{Ralph2008,Brataas2012} provide the basic mechanism to manipulate the magnetization state of MTJs in STT MRAMs \cite{Khvalkovskiy2013,Kent2015}. The main advantages of STT MRAMs, compared to other types of non-volatile RAMs, are the low power consumption, long retention times, high endurance, and a limited number of fabrication steps \cite{Kent2015,Chun2013,Hanyu2016,Apalkov2016,Wang2020,Ikegawa2020}. For these reasons, STT MRAMs are considered to be universal memories with the capabi\-li\-ty to replace flash memories, last-level cache, and static RAMs in embedded applications. MRAMs have also better downscaling properties compared to static RAMs, especially when MTJs with strong perpendicular magnetic anisotropy are used to store digital information. In STT devices, however, the spin polarization of the current flowing through the MTJ is initially aligned with the magnetization of the free layer, which results in large and stochastic incubation delays \cite{Ralph2008,Sun2000}. Moreover, the large current densities required for fast operation stress the tunnel barrier and accelerate its degradation \cite{Heindl2011}. These aspects limit the reliable switching speed to seve\-ral nanoseconds, and hence preclude the use of STT MRAMs in applications that run close to the clock speed of the central processing unit.

\begin{figure}[b]
	\includegraphics[scale=1]{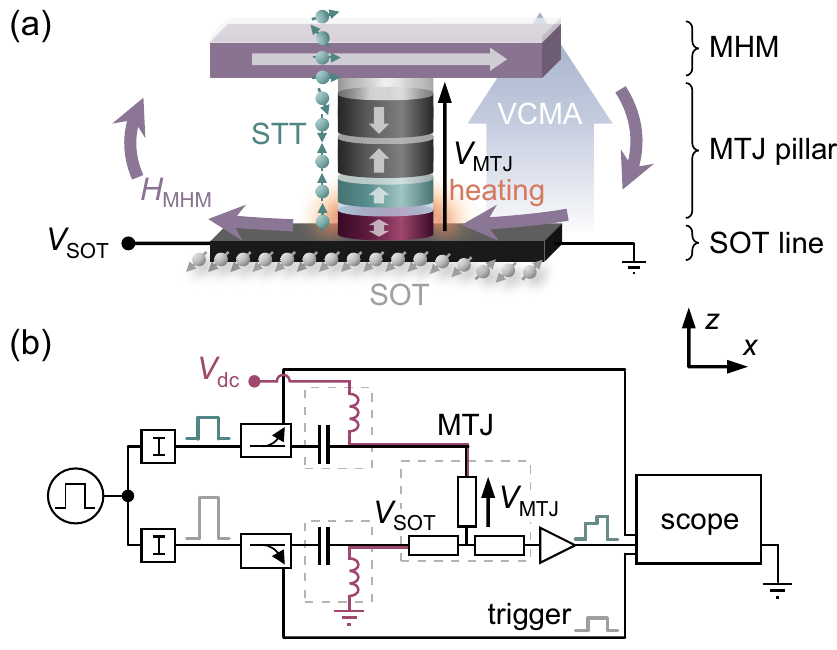}
	\caption{\label{fig1} (a) Schematics of the three-terminal device allowing for field-free SOT switching with an illustration of the effects at play when applying a bias voltage on the MTJ: STT, current-induced heating, and the VCMA effect add to the SOT driven by $V_{\text{SOT}}$. Purple arrows represent the stray field produced by the magnetic hard mask (MHM) that enables zero-field switching. (b) Simplified schematics of the circuitry used for real-time (black) and dc (purple) detection of switching induced by SOT (bottom) and MTJ bias (top) pulses. }
\end{figure}

To replace or complement STT-induced switching in MTJs, alternative mechanisms relying on the voltage control of magnetic anisotropy (VCMA) \cite{Weisheit2007,Maruyama2009} and spin-orbit torques (SOTs) \cite{Miron2011,Liu2012,Pai2012,Garello2013,Cubukcu2014,Manchon2019} have been proposed. The main advantage of the VCMA-enabled switching is low energy consumption \cite{Nozaki2010,Wang2012,Kanai2013,Kato2018,Inokuchi2017}, whereas SOTs attracted much interest for their promise of faster switching, low error rates, and endurance of the barrier \cite{Cubukcu2014,Garello2014,Dieny2020,Zhu2020,Grimaldi2020,Zhang2015,Aradhya2016,Fukami2016,Lee2016,Prenat2016,Decker2017,Rowlands2017,Cubukcu2018}. Since SOTs do not require the electric current to flow into a magnet for switching, SOT MRAMs can use an in-plane current flowing parallel to the free layer for writing, and an out-of-plane current passing through the MTJ pillar for reading. The read and write paths are thus separated in a three-terminal device geometry [Fig.~\ref{fig1}(a)], which reduces read and write errors \cite{Aradhya2016,Garello2018} while removing restrictions on the switching speed due to the high tunneling current \cite{Grimaldi2020,Krizakova2020}. In addition, the orthogonal alignment of the free-layer magnetization and the SOT allows for minimizing the delay preceding the reversal \cite{Garello2014,Lee2013,Jhuria2020}. 

Three-terminal MTJs operated by SOTs, however, also present disadvantages compared to two-terminal MTJs operated by STT. Specifically, three-terminal SOT devices have a larger footprint, consume more current for switching, and require a symmetry-breaking mechanism to define the switching polarity in perpendicularly magnetized layers \cite{Miron2011,Manchon2019}. The most straightforward solution for the latter requirement is to apply an external magnetic field parallel to the SOT-current track \cite{Miron2011,Avci2012}, which is however impractical for applications. This issue has triggered numerous efforts to achieve zero-field switching by SOTs, including the introduction of material asymmetry \cite{Yu2014a,Safeer2016}, tilted magnetic anisotropy \cite{You2015}, exchange bias \cite{Fukami2016,VanDenBrink2016,Krishnaswamy2020,Oh2016}, coupling to a reference ferromagnet \cite{Lau2016,Baek2018,Luo2019}, built-in stray field \cite{Miron2011,Krizakova2020,Garello2019}, and two-pulse schemes \cite{VanDenBrink2014,DeOrio2019,Wang2018}.

Recent measurements have shown that field-free switching by sub-ns SOT pulses can be achieved with high reliability in three-terminal MTJ devices with an embedded magnetic hard mask \cite{Krizakova2020,Garello2019}, which cannot be attained in the same device by STT because of back-hopping \cite{Sun2009,Devolder2020}. Moreover, time-resolved measurements have shown that very narrow switching time distributions can be obtained in three-terminal devices by assisting SOT switching by an increased in-plane magnetic field, VCMA, or STT \cite{Grimaldi2020,Krizakova2020}. Such measurements also evidenced that different effects play a role in determining the switching speed and efficiency of the MTJs, namely the SOT, VCMA, STT, and current-induced heating. Despite this initial work, a detailed quantitative understanding of how these different effects concur in the switching process is lacking. Optimizing their interplay is not only relevant to understand the switching dy\-na\-mics in MTJs, but also to reduce the critical SOT current required for switching.

In this paper, we present a systematic study of SOT-induced switching of three-terminal MTJs assisted by an MTJ bias applied to the magnetic pillar. The MTJ bias is responsible for inducing VCMA, STT, as well as contributing to Joule heating [Fig.~\ref{fig1}(a)]. Using a combination of real-time and post-pulse electrical measurements, we show that the simultaneous injection of dual pulses across the SOT-track and MTJ allows us to reduce the overall writing energy and voltages relative to switching by SOT or STT alone without compromising on the sub-ns writing capabilities of SOT switching \cite{Garello2014,Cubukcu2018,Jhuria2020}. We further analyze the strength, symmetry, and time scale of the different effects induced by the MTJ bias and propose a simple model that allows for separating their contributions and evaluating their impact on the SOT-induced switching of MTJs of variable size. Our results provide guidance for finding the optimal balance between SOT and STT currents and thus improving the switching efficiency, reliability, and speed of three-terminal MTJ devices. 

The paper is organized as follows. Section~\ref{2} describes the MTJs, the measurement setup and protocol, and a micromagnetic model of the devices. The effects induced by applying a voltage bias on the MTJ are first identified and evaluated by performing dc switching measurements in Sec.~\ref{3}, and then investigated systematically by time-resolved measurements at ns and sub-ns time scales. Section~\ref{4} examines the impact of the MTJ bias on the critical switching conditions. Section~\ref{5} provides experimental and micromagnetic insight into the effects of the bias on the activation delay and magnetization reversal. Section~\ref{6} presents a method to separate and evaluate these effects in relation to the geometrical and material parameters of the MTJs as well as the pulse duration.


\section{\label{2}Samples and methods}

\subsection{Devices}

The study is performed on three-terminal MTJ devices [Fig.~\ref{fig1}(a)] suitable for zero-external-field switching by SOT. This functionality is enabled by the stray field of an in-plane magnetized ferromagnetic layer that is embedded in the hard mask used to pattern the SOT injection line \cite{Krizakova2020,Garello2019}. The structure of the MTJ pillar is, from bottom to top, CoFeB(9)/MgO/CoFeB(10)/W(3)/ Co(12)/Ru(8.5)/Co(6)/Pt(8)[Co(3)/Pt(8)]$_6$Ru(50),~whe\-re the numbers in parentheses indicate the thickness of each layer in angstroms. The CoFeB reference layer is ferromagnetically coupled to the synthetic antiferromagnetic structure (SAF) by the W spacer. The pillar is grown on a $\beta$-W SOT line with a resistivity of $160\,\mu\Omega\,$cm and an effective spin Hall angle of -0.32 \cite{Garello2018}. Unless otherwise stated, the measurements have been performed on a single MTJ device with a cross-section of the SOT line of $(170 \times 3.5)\,\mbox{nm}^2$, a circular free layer with a diameter of 60\,nm, a tunnel barrier with the resistance-area product $RA = 24\,\Omega\,\mu\mbox{m}^2$, and tunnel magnetoresistance (TMR) of 104\%, corresponding to a spin polarization of 0.59 according to Julliere's model. The average stray field produced by the SAF at the position of the free layer is $H_{\text{offset}} = 150\,$Oe along -$z$, and the stray field of the magnetic hard mask is $H_{\text{MHM}} \approx 400$\,Oe along -$x$.

\subsection{Electrical measurements}

Figure~\ref{fig1}(b) shows the main components of the experimental setup, which consists of rf and dc paths for real-time and post-pulse detection of the magnetization states, respectively. The setup allows for the simultaneous application of pulsed SOT and MTJ currents, as described in detail in Ref.~\cite{Grimaldi2020}. To study the effect of a dc bias on the MTJ, a slowly varying voltage ($V_{\text{dc}}$) is applied between the top electrode and a grounded bottom electrode [along the purple path in Fig.~\ref{fig1}(b)], and the MTJ resistance ($R_{\text{MTJ}}$) is measured by modulating $V_{\text{dc}}$ at 10\,Hz with an amplitude of 20\,mV.

In the time-resolved measurements, a rectangular pulse supplied by a fast pulse generator is split in two phase-matched parts that are individually attenuated and injected to the top and bottom electrodes of the MTJ device. The pulse applied to the bottom electrode ($V_{\text{SOT}}$) generates the SOT driving the reversal of the free layer, whereas the pulse applied to the top electrode ($V_{\text{top}}$) sets a potential difference across the MTJ ($V_{\text{MTJ}}$). From an equivalent circuit model of the device, it can be shown that $V_{\text{MTJ}} = \left(V_{\text{top}} - \frac{V_{\text{SOT}}}{2}\right)/\left(1 + \frac{R_{\text{SOT}}}{4R_{\text{MTJ}}}\right)$, where $R_{\text{SOT}}$ is the resistance of the SOT-line. Therefore, to perform an SOT-only switching experiment, we apply a compensating voltage pulse $V_{\text{top}}$, because $V_{\text{SOT}}$ itself generates a finite $V_{\text{MTJ}}$. Controlling the value of $V_{\text{MTJ}}$ by $V_{\text{top}}$ allows us to investigate the effects arising from the MTJ bias in real-time. Note that the overall SOT current can be considered independent of $V_{\text{MTJ}}$ because the increase and decrease of the current driven by $V_{\text{MTJ}}$ on one and the other side of the SOT-line cancel out. In the time-resolved study (Sec.~\ref{5}-\ref{6}), we set $V_{\text{SOT}}= 1$--$1.5V_{\text{c}}$, where $V_{\text{c}}$ is the critical voltage that results in a 50\% probability of switching at $V_{\text{MTJ}} = 0$ for 15-ns-long pulses, and we vary the MTJ bias within the range $|V_{\text{MTJ}}| \leq 3.2|V_{\text{SOT}}|$. In these conditions, the switching outcome is determined by SOT, consistently with $V_{\text{MTJ}}$ being lower than the threshold voltage for STT switching. The latter is more than four times higher compared to $V_{\text{c}}$, such that at $V_{\text{MTJ}} = 3.2V_{\text{SOT}}$ we reach only 60\% (70\%) of the critical STT voltage for P-AP (AP-P) switching, where P and AP denote the parallel and anti-parallel state of the MTJ, respectively. In the following, we quantify the MTJ bias using the ratio $V_{\text{MTJ}}/V_{\text{SOT}}$.

To achieve $V_{\text{MTJ}} < -0.5V_{\text{SOT}}$, a balun transformer is used in the setup (instead of a standard power divider) to reverse the polarity of $V_{\text{top}}$. In this way, a single pulse generator can simultaneously supply $V_{\text{SOT}}$ and $V_{\text{MTJ}}$ with opposite polarities. The voltage pulse transmitted through the device is acquired on a sampling oscilloscope and normalized to the voltage difference between the P and AP states of the MTJ \cite{Cui2010,Sampaio2013,Hahn2016}. The normalized voltage time trace ($V_{\text{sw}}$) tracks the TMR during the pulse injection. To observe the voltage difference between the two states, a finite current tunneling through the MTJ is needed, hence the real-time detection of magnetization reversal requires the application of small $V_{\text{MTJ}} \neq 0$, which is not necessary in post-pulse measurements of $R_{\text{MTJ}}$. Additionally, the state before and after each pulse is verified by a post-pulse measurement of $R_{\text{MTJ}}$ with $V_{\text{dc}} = 0$. 

\begin{figure}
	\includegraphics[scale=1]{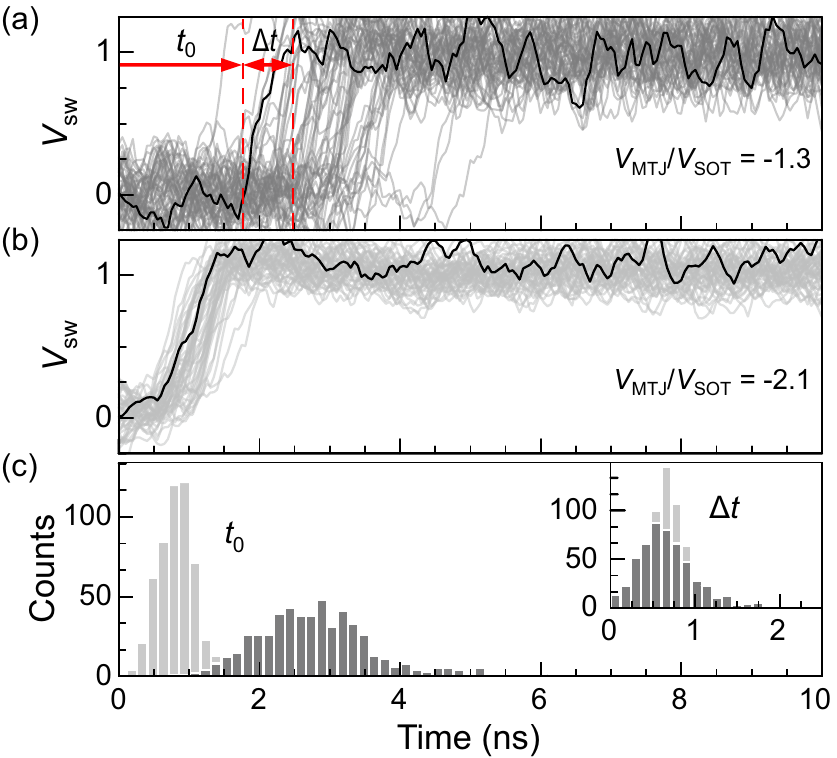}
	\caption{\label{fig2} Representative voltage time traces ($V_{\text{sw}}$) recorded during AP-P reversal, smoothed with a 0.5\,ns window, at $V_{\text{SOT}} = 0.27$\,V with (a) $V_{\text{MTJ}} = -1.3V_{\text{SOT}}$ and (b) $V_{\text{MTJ}} = -2.1V_{\text{SOT}}$. The black lines are examples of individual time traces. (c) Statistical distributions of the activation delay ($t_0$) and transition time ($\Delta t$) derived from the $V_{\text{sw}}$ traces fitted to a piece-wise linear function. The distributions are compiled from 500 switching events, 50 of which are displayed in (a,b). }
\end{figure}

Figure~\ref{fig2} shows an overlay of typical single-shot measurements of AP-P switching induced by SOT in the presence of $V_{\text{MTJ}} \approx 28\%$ [Fig.~\ref{fig2}(a)] and 46\% [Fig.~\ref{fig2}(b)] of the critical STT voltage. Each $V_{\text{sw}}$ comprises a single ‘jump’ stochastically delayed from the pulse onset. To quantify and compare these switching signatures in various conditions, we fit each trace by a piecewise linear function and derive its breakpoints [Fig.~\ref{fig2}(a)]. In line with the expectation that switching occurs via nucleation and propagation of a single domain-wall \cite{Grimaldi2020,Mikuszeit2015,Baumgartner2017}, we refer to the initial delay as the activation delay ($t_0$) and to the duration of the magnetization reversal as the transition time ($\Delta t$). Figure~\ref{fig2}(c) shows the distributions of $t_0$ and $\Delta t$ compiled from 500 switching events. Clearly, increasing $V_{\text{MTJ}}$ strongly reduces both $t_0$ and $\Delta t$, in agreement with previous results \cite{Grimaldi2020,Krizakova2020}.

\subsection{Micromagnetic simulations}

We used MuMax$^3$ \cite{Vansteenkiste2014} to simulate the current-induced magnetization reversal in an MTJ by simultaneous action of SOT and MTJ bias. For simulation purposes, the MTJ structure was modeled by a single magnetic disk with a diameter of 60\,nm and a thickness $t_{\text{FL}} = 0.9$\,nm representing the CoFeB free layer. The saturation magnetization $M_{\text{S}} = 1.1$\,MA/m, perpendicular magnetic anisotropy $K_{\text{u}} = 845$\,kJ/m$^3$, exchange stiffness $A_{\text{ex}} = 15$\,pJ/m, and the strength of interfacial Dzyaloshinskii-Moriya interaction $D = 0.15$\,mJ/m$^2$ were selected according to measurements of  $M_{\text{S}}$ and $K_{\text{u}}$ performed at room temperature and literature values for $A_{\text{ex}}$. The Gilbert damping parameter was set to 0.2. The free layer was placed into a homogeneous magnetic field of 400\,Oe in the -$x$ direction and 40\,Oe in the -$z$ direction to model the effects of $H_{\text{MHM}}$ and $H_{\text{offset}}$, respectively. The SOT and MTJ bias were supplied from rectangular pulses with a variable length of 0.8 to 20\,ns and rising/falling edges of about 0.15\,ns following a Gauss error function profile. The SOTs were modeled by assuming a homogeneous current density $|j_{\text{SOT}}| = 72$--$100$\,MA/cm$^2$, and an effective spin Hall angle of -0.3 for the dampinglike component, whereas the fieldlike SOT, which is low in W/CoFeB bilayers \cite{Garello2018} and has a small effect on the switching times in our case, was neglected. The dampinglike STT was modeled by a current $|j_{\text{MTJ}}| = 0$--3.6 (0--1.8)\,MA/cm$^2$ for the reversal from the P (AP) state, equivalent to $|V_{\text{MTJ}}| \approx 0-0.9$\,V ($|V_{\text{MTJ}}/V_{\text{SOT}}| \leq 3$) for a tunnel magnetoresistance of 104\% (corresponding to spin polarization of 0.59).

The thermal effects were taken into account by scaling $M_{\text{S}}$, $K_{\text{u}}$, and  $A_{\text{ex}}$ in time due to Joule heating induced by the current passing through the SOT-track and MTJ, according to the model proposed in Ref.~\cite{Grimaldi2020}. A stochastic thermal field was not included in the simulations. First, a current-induced temperature ($T$) change in the free layer was calculated by finite-element analysis using \textsc{comsol} as a function of time. Then, the temporal dependence $T(t)$ was approximated by a double-exponential rise and decay function scaled by the currents through the SOT-line and the MTJ, and inserted into the established scaling models \cite{Moreno2016,Lee2017b}: $M_{\text{S}}(T) = M_{\text{S}}(0)\left(1 - \frac{T}{T_{\text{C}}}\right)^b$, $K_{\text{u}}(T) = K_{\text{u}}(0)\left(\frac{M_{\text{S}}(T)}{M_{\text{S}}(0)}\right)^p$, and $A_{\text{ex}}(T) = A_{\text{ex}}(0)\left(\frac{M_{\text{S}}(T)}{M_{\text{S}}(0)}\right)^q$, where $T_{\text{C}} = 850$\,K, $b = 1$, $p = 2.5$, $q = 1.8$. The VCMA effect was considered as: $K_{\text{u}}(V_{\text{MTJ}},T) = K_{\text{u}}(T) - \frac{RA \varepsilon }{t_{\text{MgO}} t_{\text{FL}}}j_{\text{MTJ}}$, assuming the VCMA coefficient $\varepsilon = 30$\,fJ/V\,m and a thickness of the tunnel barrier $t_{\text{MgO}} = 1$\,nm.

The magnetization dynamics was solved in a simulation box of $(64 \times 64 \times 0.9)$\,nm$^3$ with $(2 \times 2)$\,nm$^2$ discretization in the plane. Before the pulse onset, the magnetization was initialized up (down) for P-AP (AP-P) switching to simulate the reversal for a reference layer magnetized up, and let relax in the external field. The simulated $t_0$ and $\Delta t$ were obtained from the temporal evolution of the normalized $z$ component of the magnetization averaged over the free layer.

\section{\label{3}Concurrent effects due to MTJ bias on field-induced reversal}

\begin{figure}
	\includegraphics[scale=1]{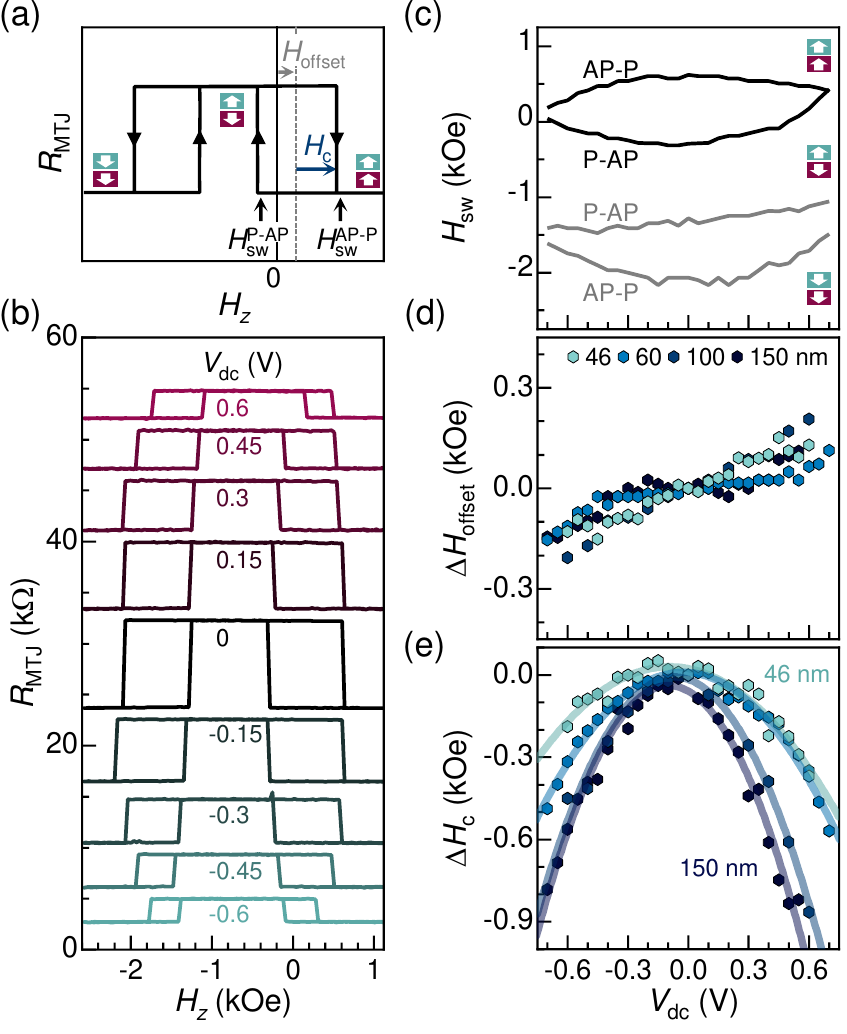}
	\caption{\label{fig3} Field-induced switching assisted by a dc voltage bias. (a) Schematics of an $R$-$H$ loop showing the switching fields ($H_{\text{sw}}$), the offset field ($H_{\text{offset}}$), and the coercive field ($H_{\text{c}}$). The block arrows show the direction of the magnetization of the free and reference layers at different fields. (b) $R$-$H$ loops measured at different $V_{\text{dc}}$ applied to the MTJ with a diameter of 60\,nm. The measurements are offset along the $y$-axis for clarity. (c) $H_{\text{sw}}$ corresponding to P-AP and AP-P reversals of the free (black lines) and reference (gray lines) layer extracted from (b). (d) Variation of the offset field $\Delta H_{\text{offset}} = H_{\text{offset}}(V_{\text{dc}}) - H_{\text{offset}}(0)$ and (e) coercive field $\Delta H_{\text{c}} = H_{\text{c}}(V_{\text{dc}}) - H_{\text{c}}(0)$ as a function of $V_{\text{dc}}$ and MTJ diameter. Solid lines in (e) are fits to the data (see text).}
\end{figure}

We first classify the different effects originating from the MTJ bias on our devices. For clarity, we start by exemplifying the effects at play in field-induced magnetization reversal using a dc bias. In this experiment, $V_{\text{dc}}$ is applied to the top electrode of the MTJ, the bottom electrode is grounded, and the MTJ resistance is measured during a sweep of the external magnetic field along $z$. A typical $R$-$H$ loop is sketched in Fig.~\ref{fig3}(a). The impact of $V_{\text{dc}}$ on the field-induced switching of the free and reference layers is visible in Fig.~\ref{fig3}(b). The switching of the free layer takes place close to $H_z = 0$, whereas the switching of the reference layer is offset by $\approx$1.8\,kOe, the pinning field from the SAF. The TMR \cite{Li2004} as well as the switching fields ($H_{\text{sw}}$) of both layers depend on the $V_{\text{dc}}$ bias. The values of $H_{\text{sw}}$ extracted from the hysteresis loops are plotted in Fig.~\ref{fig3}(c), their dependence on $V_{\text{dc}}$ can be described as follows: i) $|H_{\text{sw}}|$ of the free and reference layer reduce with increasing $|V_{\text{dc}}|$, ii) the loops shift towards the positive (negative) direction for $V_{\text{dc}} > 0$ ($V_{\text{dc}} < 0$), and iii) the loops of the free layer are narrowed more by $V_{\text{dc}} > 0$ than by $V_{\text{dc}} < 0$, whereas the opposite trend is found for the reference layer. As we will show later, these three effects can be ascribed to current-induced heating in the MTJ pillar, STT, and VCMA, respectively, which all emerge from the potential difference across the junction. Such effects can be disentangled and quantified by analyzing their dependence on $V_{\text{dc}}$. In the following, we will focus on the switching of the free layer, although the same reasoning can be applied to the reference layer.

In line with previous studies \cite{Chavent2016,Mihajlovic2020}, we define the offset field $H_{\text{offset}} = \left(H_{\text{sw}}^{\text{P-AP}} + H_{\text{sw}}^{\text{AP-P}} + \Delta H_{\Delta T} \right)/2$ and the coercive field $H_{\text{c}} = \left(H_{\text{sw}}^{\text{P-AP}} - H_{\text{sw}}^{\text{AP-P}} + \Delta H_{\Delta T}\right)/2$ in terms of the sum and difference of $H_{\text{sw}}$ for different switching polarities, as illustrated in Fig.~\ref{fig3}(a). We include in these definitions the change of the switching field $\Delta H_{\Delta T}$, equal to the difference between $[\Delta H_{\text{sw}}(V_{\text{dc}} > 0) + \Delta H_{\text{sw}}(V_{\text{dc}} < 0)]/2$ for P-AP and AP-P reversal, which accounts for the different amount of current-induced heating under the same $|V_{\text{dc}}|$ when switching from the low or high resistive state. At $V_{\text{dc}} = 0$, $H_{\text{offset}}$ is given solely by the stray field from the SAF. At finite $V_{\text{dc}}$, the current tunneling through the MTJ gives rise to STT. In general, the current driven by $V_{\text{dc}} > 0$ promotes the AP state, whereas the current due to $V_{\text{dc}} < 0$ promotes the P state of the MTJ. The generated STT is proportional to the tunneling current, and consequently to the positive shift of $H_{\text{offset}}$ with increasing $V_{\text{dc}}$ \cite{Sun2000}. This approximate linear dependence is demonstrated in Fig.~\ref{fig3}(d) on four different MTJ devices with a diameter of 50--150\,nm. Even though we observe a weak device-to-device variation, the slope $H_{\text{offset}}/V_{\text{dc}}$ does not scale with the size of the MTJ. The average slope of $\approx$0.2\,kOe/V, i.e. 7.1\,kOe per A/$\mu$m$^2$, can be understood as the STT efficiency of our devices, which is close to that reported for similar devices in Ref.~\cite{Mihajlovic2020}.

Figure~\ref{fig3}(e) shows the reduction of $H_{\text{c}}$ due to $V_{\text{dc}}$ in the four MTJ devices. Most significant is the stronger decrease of $H_{\text{c}}$ with $V_{\text{dc}}$ in the larger devices. Such sca\-ling can be explained by the reduction of the magnetic anisotropy of the free layer due to the current-induced temperature increase in the device. The equilibrium temperature is given by the balance between Joule heating $\propto {V_{\text{dc}}}^2/R_{\text{MTJ}}$, (where $R_{\text{MTJ}}$ is inversely proportional to the pillar cross-section) and heat dissipation through the pillar cross-section and side walls. Both these terms increase with the MTJ diameter; however, the dissipation through the side walls becomes relatively more important in the smaller devices, reducing the equilibrium temperature compared to larger devices. In addition to heating, the magnetic anisotropy of the layers adjacent to the tunnel barrier can be modified by the accumulation of electric charges at the barrier’s interfaces induced by $V_{\text{dc}}$, that is, by the VCMA effect \cite{Weisheit2007,Maruyama2009}. In the CoFeB/MgO/CoFeB system, the VCMA weakens (strengthens) the anisotropy of the bottom free layer and strengthens (weakens) the anisotropy in the top reference layer when a positive (negative) bias is applied to the top electrode of the MTJ \cite{Wang2012}. To distinguish between the two mechanisms affecting the magnetic anisotropy, we fit the change of $H_{\text{c}}$ in Fig.~\ref{fig3}(e) to the function $\Delta H_{\text{c}} = a_1 V_{\text{dc}}$ + $a_2 {V_{\text{dc}}}^2$, where the linear and the quadratic term account for the VCMA and thermal effects, respectively. As the MTJ diameter increases from 50 to 150\,nm, we find that $a_2$ increases from 0.8\,kOe/V$^2$ to 2.1\,kOe/V$^2$, confirming the larger role played by Joule heating in the larger MTJs. On the other hand, $a_1 \approx 0.24$--0.4\,kOe/V has no significant size dependence. Assuming the VCMA coefficient $\varepsilon =a_1 M_{\text{S}} t_{\text{FL}} t_{\text{MgO}}/2$, where $M_{\text{S}} = 1.1$\,MA/m, $t_{\text{FL}} = 0.9$\,nm, and $t_{\text{MgO}} = 1$\,nm, we obtain $\varepsilon = $12--20\,fJ/V\,m, which is in agreement with li\-tera\-ture values for this system \cite{Maruyama2009,Wang2012,Grezes2017,Wu2021}. However, $\varepsilon$ is lower than our former estimate \cite{Grimaldi2020} that did not take the thermal effects into account, resulting in an overestimation of the VCMA effect.

These measurements allowed us to identify the effects induced by the MTJ bias, and quantify the extent to which they affect the field-induced switching in the dc limit. In Sec.~\ref{5}-\ref{6}, we will show how these effects relate to SOT switching and how they unfold in time.


\section{\label{4}Critical conditions for switching by SOT and STT}

\begin{figure}[b]
	\includegraphics[scale=1]{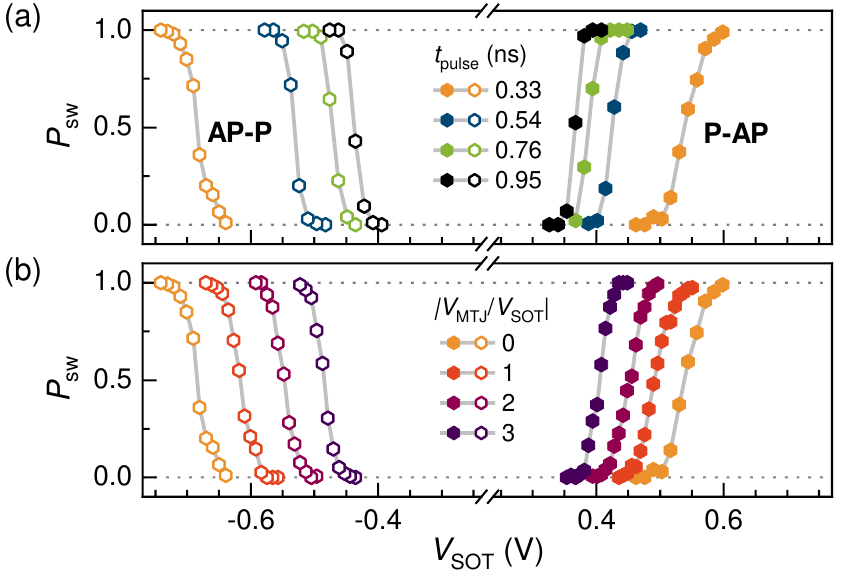}
	\caption{\label{fig4} Probability of AP-P (open symbols) and P-AP (full symbols) switching measured post-pulse as a function of $V_{\text{SOT}}$. (a) Switching at $V_{\text{MTJ}} = 0$ for different pulse widths and (b) switching at different $V_{\text{MTJ}}$ for $t_{\text{pulse}} = 0.33$\,ns.}
\end{figure}

\begin{figure*}
	\includegraphics[scale=1]{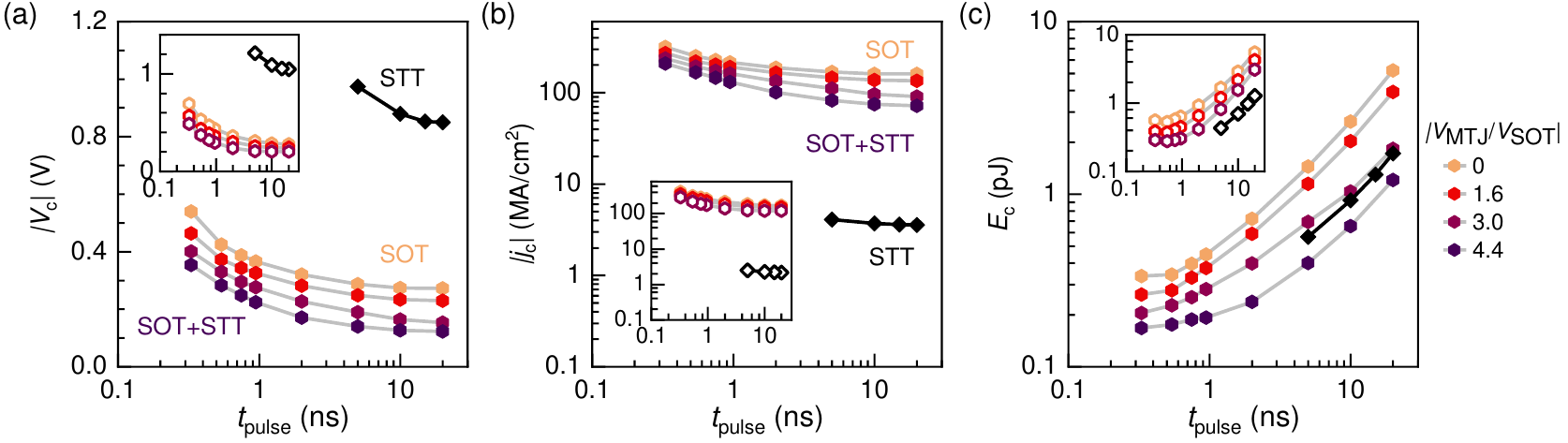}
	\caption{\label{fig5} Dependence of the critical switching parameters for STT (black diamonds) and SOT (colored hexagons). The purple data set, $|V_{\text{MTJ}}/V_{\text{SOT}}| = 4.4$, represents the borderline between SOT- and STT dominated switching. The main panels refer to P-AP switching, the insets to AP-P switching.  (a) Critical voltage as a function of $t_{\text{pulse}}$. $V_{\text{c}}$ corresponds to the pulse amplitude for which $P_{\text{sw}} = 0.5$, $V_{\text{SOT}}$ for the SOT-dominated switching, and $V_{\text{MTJ}}$ for STT switching. (b) Critical current density $j_{\text{c}}$ and (c) critical energy $E_{\text{c}}$ as a function of $t_{\text{pulse}}$. }
\end{figure*}

Knowledge of the critical conditions for switching induced by SOT, STT, and their combination at ns and sub-ns time scale is required to understand the pheno\-me\-na at play. Therefore, we studied the probability of switching ($P_{\text{sw}}$) of the same 60\,nm MTJ device as a function of $V_{\text{SOT}}$ and $V_{\text{MTJ}}$ by measuring its resistance after each writing pulse. Figure~\ref{fig4}(a) shows the evolution of $P_{\text{sw}}$ with $V_{\text{SOT}}$ in a range close to the switching threshold for different widths ($t_{\text{pulse}}$) of the SOT pulses, defined as full width at half maximum. In the intrinsic switching regime, which corresponds to the sub-ns time scale, $V_{\text{c}} \propto 1/t_{\text{pulse}}$ \cite{Garello2014,Krizakova2020}. Therefore, $V_{\text{SOT}}$ strongly increases with decreasing $t_{\text{pulse}}$. Considering AP-P reversal, a threefold reduction of $t_{\text{pulse}}$ from 1\,ns to 0.33\,ns requires an increase of $V_{\text{SOT}}$ from 0.45\,V to 0.7\,V. However, by extrapolation, a tenfold reduction of $t_{\text{pulse}}$ down to 100\,ps would require an increase of $V_{\text{SOT}}$ up to 1.6\,V, equivalent to 5\,mA. This current represents a technological barrier because it is well above the typical current ($<$100\,$\mu$A) required for embedded memory. However, for a given $t_{\text{pulse}}$, $V_{\text{SOT}}$ can be decreased by downscaling the size of the MTJ as well as by applying a finite $V_{\text{MTJ}}$ during the SOT pulse, as shown in Fig.~\ref{fig4}(b). In the selected range of $V_{\text{MTJ}} \leq 3|V_{\text{SOT}}|$, the switching outcome is determined by the sign of $V_{\text{SOT}}$ and $H_{\text{MHM}}$, while $V_{\text{MTJ}}$ facilitates the reversal.

In principle, SOT switching is symmetric with respect to $V_{\text{SOT}} \gtrless 0$ \cite{Miron2011}. However, the stray field from the SAF is not balanced in our devices, giving rise to $H_{\text{offset}}$ that stabilizes the AP state at the expense of the P state. As a result, $V_{\text{c}}$ is lower for P-AP switching compared to AP-P [Fig.~\ref{fig4}(a)]. Further, $V_{\text{MTJ}}$ can induce an additional asymmetry of $P_{\text{sw}}$ \textit{vs} $V_{\text{SOT}}$, because STT and VCMA can each either assist or hinder the SOT switching for a given polarity of $V_{\text{SOT}}$. In this example, we selected $V_{\text{MTJ}}$ to be positive, such that the switching in both configurations is assisted by VCMA. STT counteracts one of the reversal orientations, but its effect for $t_{\text{pulse}} < 1$\,ns can be neglected, as we will show in Section~\ref{6}.

To characterize the probability of STT switching in the same device, we measured $P_{\text{sw}}$ as a function of $V_{\text{MTJ}}$ with the SOT-input electrode terminated by $50\,\Omega$, so that $V_{\text{SOT}} = 0$. Figure~\ref{fig5}(a) shows that the critical switching voltage is significantly higher for STT compared to SOT, and that their difference becomes even more pronounced when a moderate $V_{\text{MTJ}}$ is applied to the MTJ to assist the SOT-induced switching. However, Fig.~\ref{fig5}(b) shows that the critical current density $j_{\text{c}}$ (recalculated from $V_{\text{c}}$) ranges between 3--4\,MA/cm$^2$ for STT, whereas it reaches 160--180\,MA/cm$^2$ for SOT, considering pulses with the same duration. As a result, STT surpasses SOT in terms of energy efficiency, as shown by calculating the critical energy $E_{\text{c}} = ({V_{\text{SOT}}}^2/R_{\text{SOT}} + {V_{\text{MTJ}}}^2/R_{\text{MTJ}})t_{\text{pulse}}$ [Fig.~\ref{fig5}(c)]. Nevertheless, back-hopping limits the STT switching to $t_{\text{pulse}} \geq 5$\,ns in our devices, whereas SOT works reliably down to sub-ns time scales. Therefore, owing to the favorable scaling of $E_{\text{c}}$ with $t_{\text{pulse}}$, SOT outperforms STT at shorter time scales. More importantly, we observe that the combination of $V_{\text{SOT}}$ and $V_{\text{MTJ}}$ leads to a reduction of $E_{\text{c}}$ below the STT threshold for all $t_{\text{pulse}}$, while at the same time preserving the capability to switch in the sub-ns regime, as demonstrated by the purple data set in Fig.~\ref{fig5}(c). These promising features of SOT switching assisted by $V_{\text{MTJ}}$ call for a closer exa\-mi\-nation of the role played by the bias during single-shot magnetization reversal events, which we present in the next section.


\section{\label{5}Dependence of the switching time on MTJ bias}

\subsection{Concurrent effects due to $V_{\text{MTJ}}$ on SOT-induced magnetization reversal}

In this section, we focus on the impact of $V_{\text{MTJ}}$ on the real-time magnetization dynamics driven by SOT. Based on the dc measurements presented in Sec.~\ref{3}, we identified VCMA, STT, and Joule heating as the main consequences of $V_{\text{MTJ}}$. These effects differ in the extent to which they contribute to the switching and in their symmetry with respect to SOT. In our experiments, STT promotes the P (AP) state when $V_{\text{MTJ}} < 0\,(V_{\text{MTJ}} > 0)$, the VCMA effect weakens (strengthens) the anisotropy of the free layer when $V_{\text{MTJ}} > 0\,(V_{\text{MTJ}} < 0)$, and Joule heating lowers the energy barrier for magnetization reversal regardless of the polarity of $V_{\text{MTJ}}$. With this knowledge, we now concentrate on their impact in determining the speed of single-shot SOT switching events.

\begin{figure}
	\includegraphics[scale=1]{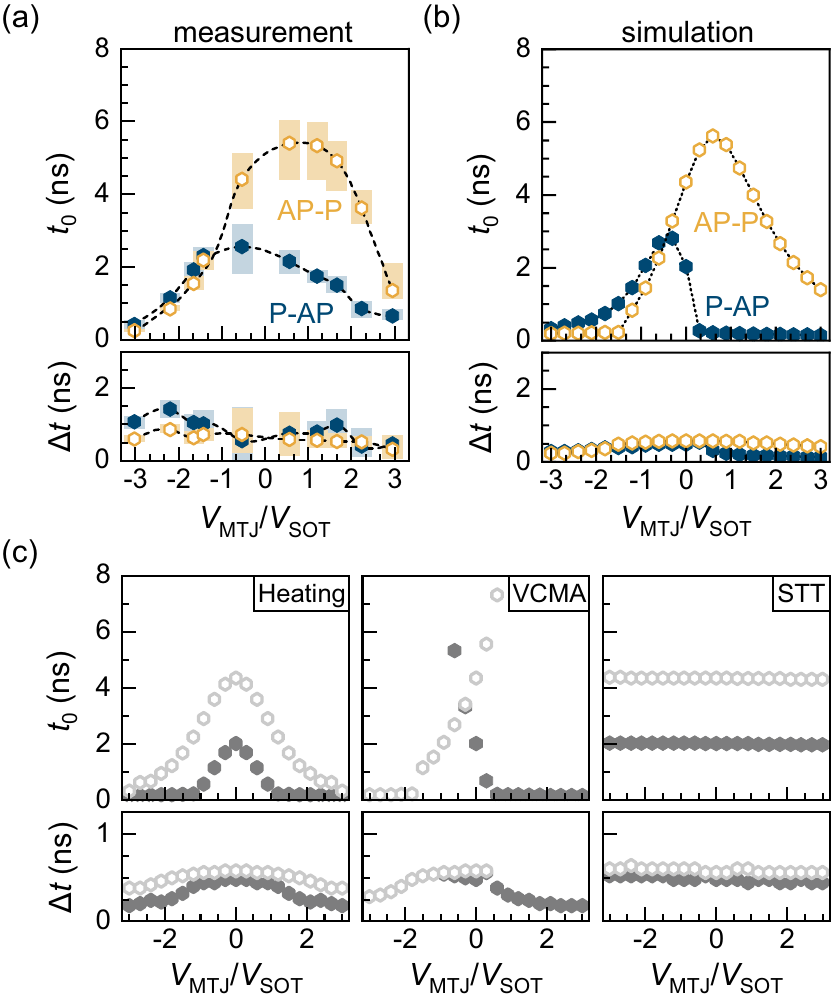}
	\caption{\label{fig6} (a) Median (symbols) and interquartile range (shaded areas) of the statistical distributions of $t_0$ and $\Delta t$ as a function of $V_{\text{MTJ}}/V_{\text{SOT}}$. The data have been obtained by measuring 500 single-shot switching traces for each point using a fixed pulse amplitude $|V_{\text{SOT}}| = 305$\,mV, which is equivalent to 1.1$V_{\text{c}}$ for AP-P switching at $V_{\text{MTJ}} = 0$. The dashed lines are a guide to the eye. (b) $t_0$ and $\Delta t$ obtained from micromagnetic simulations including the combined impact of SOT and MTJ bias. The SOT pulse amplitude is defined by a current density of 81\,MA/cm$^2$. (c) $t_0$ and $\Delta t$ obtained from the simulations by considering only one of the three effects of $V_{\text{MTJ}}$ in each plot. Full (open) symbols correspond to the P-AP (AP-P) reversal.}
\end{figure}

We compare P-AP and AP-P reversals induced by $V_{\text{SOT}} > 0$ and $V_{\text{SOT}} < 0$, respectively, acquired at fixed $|V_{\text{SOT}}|$ and different $V_{\text{MTJ}}$. Figure~\ref{fig6}(a) summarizes the statistical distribution of $t_0$ and $\Delta t$ measured over 500 switching events for each $V_{\text{MTJ}}/V_{\text{SOT}}$ ratio. The selected pulse amplitude, $|V_{\text{SOT}}| = 305$\,mV, is the lowest one required to achieve switching in all trials using 15-ns-long pulses for all values of $V_{\text{MTJ}}$ considered here. The most distinct feature is the different scaling of $t_0$ and $\Delta t$ with $V_{\text{MTJ}}$. Whereas $t_0$ changes by almost one order of magnitude in the range $|V_{\text{MTJ}}/V_{\text{SOT}}| < 3$, $\Delta t$ shows minor variations that are well within the spread of its statistical distribution. This result emphasizes the different phy\-si\-cal origins of $t_0$ and $\Delta t$, where $t_0$ is identified with the time required to nucleate a reversed domain and $\Delta t$ with the time required for its expansion across the free layer \cite{Mikuszeit2015}. The clearest feature in the $t_0$ dependence is the reduction with $|V_{\text{MTJ}}|$, which we attribute to the temperature increase. On closer inspection, two more effects are visible in Fig.~\ref{fig6}(a): i) the onset of the P-AP reversal is faster than that of the AP-P reversals in most conditions, and ii) the maxima of $t_0$ of the two switching configurations are shifted with respect to each other and $V_{\text{MTJ}} = 0$. We ascribe i) to $H_{\text{offset}}$, which stabilizes the AP state at the expense of the P state, favoring the nucleation of a domain when switching from the P state. Further, we assign ii) to VCMA, which has the opposite effect on P-AP and AP-P reversals because VCMA assists the switching when $V_{\text{MTJ}} > 0$. On the contrary, the state favored by STT depends on the sign of $V_{\text{MTJ}}$, which cannot explain the shift of $t_0$ toward  $V_{\text{MTJ}} < 0$. Both $H_{\text{offset}}$ and VCMA have little influence on $\Delta t$, which implies that they do not alter the SOT-induced domain-wall motion in a noticeable way.

\begin{figure}[b]
	\includegraphics[scale=1]{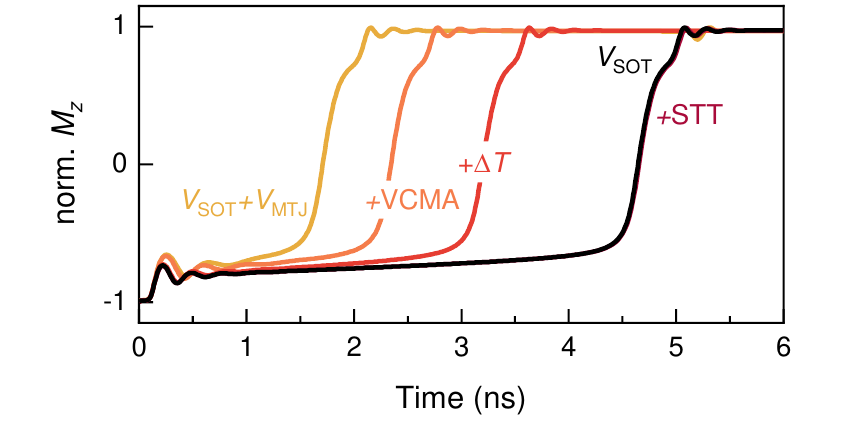}
	\caption{\label{fig7}  Simulated switching traces for AP-P reversal induced by $j_{\text{SOT}} = 81$\,MA/cm$^2$ without (black line) and with (yellow line) MTJ bias given by $V_{\text{MTJ}}/V_{\text{SOT}} = -0.9$, and taking into account only one bias-induced effect at time (orange-red lines).}
\end{figure}

To corroborate our interpretation of the data, we performed a micromagnetic study of the switching by concurrent SOT and MTJ bias pulses. In the model, we took into account Joule heating due to the current through the SOT-line, as well as through the MTJ, the VCMA effect, and STT. We selected the SOT current density ($j_{\text{SOT}}$), such that its impact is equivalent to that of 1.1$V_{\text{c}}$, and varied the MTJ bias similar to the experiment (see Sec.~\ref{2}.C for a detailed description of the simulations.) Figure~\ref{fig6}(b) shows $t_0$ and $\Delta t$ obtained from the simulated time traces, showing a good qualitative agreement with the experimental data set. The simulations reproduce the main features of the $t_0$ dependence, including the shape and the asymmetry with respect to $V_{\text{MTJ}} = 0$, as well as the weak variation of $\Delta t$ with MTJ bias. The reduction of $t_0$ is steeper for P-AP reversal compared to AP-P because the same $V_{\text{MTJ}}$ induces a higher $j_{\text{MTJ}}$ when the MTJ is in the low-resistive state. The simulated $t_0$ reaches an apparent minimum at a lower $|V_{\text{MTJ}}|$ compared to the experiment, which we attribute to the coherent oscillations of the magnetization visible in Fig.~\ref{fig7} in the first $\approx$0.5\,ns after the pulse onset, that help to overcome the domain nucleation barrier. To visualize the impact of each effect of $V_{\text{MTJ}}$ on the characteristic times, we performed a series of simulations, in which these effects were taken one at a time. The switching results due to the concurrent action of SOT and heating in the MTJ, SOT and VCMA, as well as SOT and STT are shown in Fig.~\ref{fig6}(c). The simulations confirm the symmetry of the different effects deduced from the experimental data, as well as the stronger impact of all the effects on $t_0$ compared to $\Delta t$. Figure~\ref{fig7} further exemplifies the separate influence of VCMA, heat, and STT on simulated switching. In general, Joule heating and VCMA have a larger impact on the switching dynamics compared to STT in the investigated bias range. Moreover, we find that not only heating induced by $j_{\text{SOT}}$ \cite{Krizakova2020}, but also $j_{\text{MTJ}}$-induced heating has a significant influence on $t_0$ despite $j_{\text{MTJ}}$ being two orders of magnitude smaller than $j_{\text{SOT}}$, which we ascribe to its localized effect on the free layer. Although small, we also observe a linear dependence of $\Delta t$ on STT. This variation can explain the weak trend of $\Delta t$ as a function of $V_{\text{MTJ}}$ seen in Fig.~\ref{fig6}(a). It also corro\-bo\-rates our observation \cite{Grimaldi2020} that STT can increase or decrease the domain wall velocity, i.e. reduce or prolong $\Delta t$, compared to the pure SOT reversal, despite having a minimal impact on the nucleation of a reversed domain.

\subsection{$V_{\text{SOT}}$ \textit{vs} $V_{\text{MTJ}}$ switching diagrams}

\begin{figure*}
	\includegraphics[scale=1]{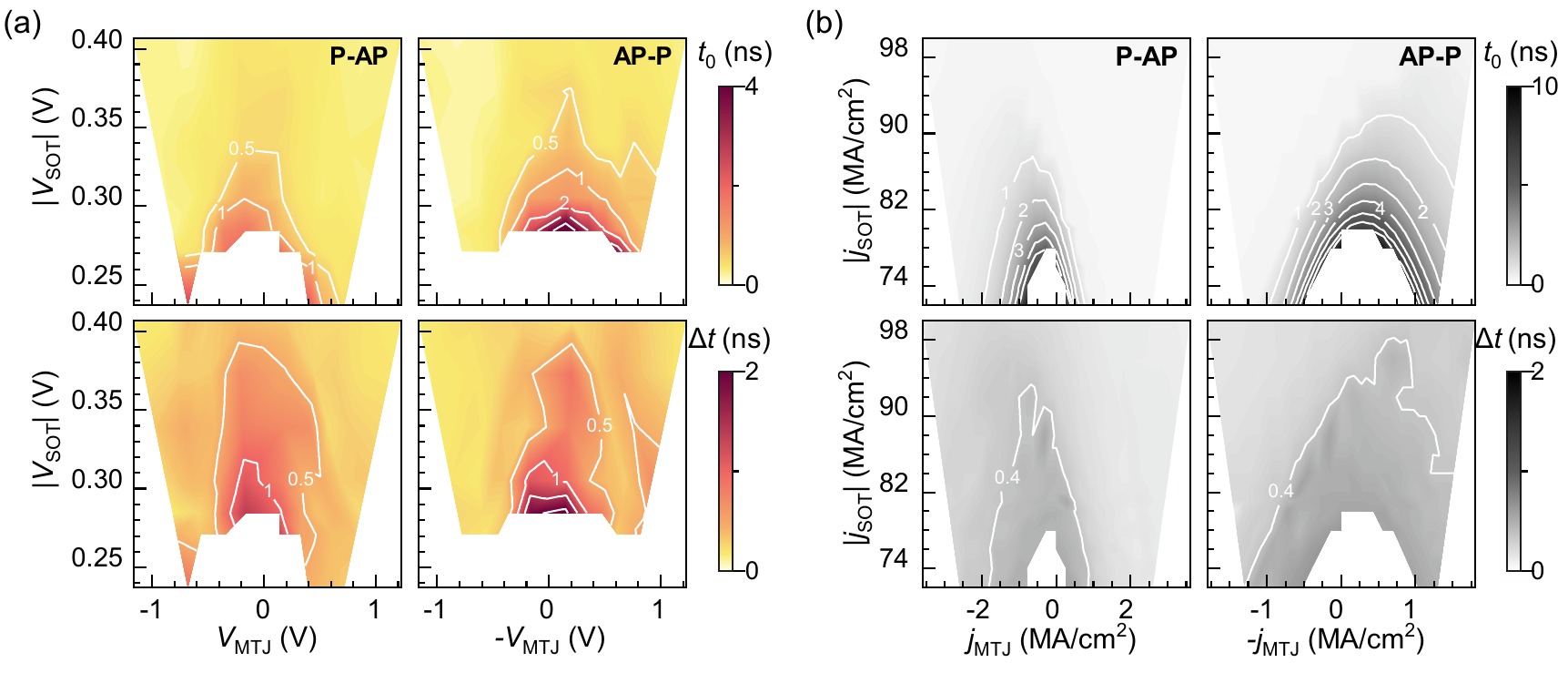}
	\caption{\label{fig8}  (a) Median value of $t_0$ (top) and $\Delta t$ (bottom) obtained from single-shot measurements of SOT-induced switching as a function of $|V_{\text{SOT}}|$ and $V_{\text{MTJ}}$. P-AP (AP-P) switching is induced by positive (negative) $V_{\text{SOT}}$. (b) $t_0$ and $\Delta t$ derived from micromagnetic simulations of the switching by the combined impact of currents flowing through the SOT-line and MTJ. The shading represents $t_0$ or $\Delta t$ in ns, according to the scale on the right. The white lines are isocurves of $t_0$ and $\Delta t$. The white areas at the bottom of the diagrams delimit the under-critical switching conditions.}
\end{figure*}

The characteristic switching times $t_0$ and $\Delta t$ can be strongly decreased by increasing the SOT and/or the MTJ bias beyond the critical threshold. Figure~\ref{fig8}(a) shows the diagrams of $t_0$ and $\Delta t$ as a function of $V_{\text{SOT}}$ and $V_{\text{MTJ}}$. The color map in the four panels represents the median value of $t_0$ (top) and $\Delta t$ (bottom) of P-AP (left) and AP-P (right) switching obtained by interpolating $10 \times 7$ data sets, each comprising 500 successful switching events. The white areas at the bottom of the plots correspond to under-critical or close-to-critical conditions (error rate $> 4 \times 10^{-2}$), where switching times were not evaluated.

The most pronounced effect is the symmetric reduction of the median value of $t_0$ with $|V_{\text{MTJ}}|$ and $|V_{\text{SOT}}|$, which is the result of the stronger SOT as well as heating in the MTJ and the SOT line. When $V_{\text{MTJ}}$ is increased by 0.7\,V from 0, or equivalently when $V_{\text{SOT}}$ is increased by 0.1\,V from $V_{\text{c}}$, the median $t_0$ decreases by a factor of 4. Therefore, in the investigated bias range, $V_{\text{SOT}}$ is more efficient in increasing the device temperature than $V_{\text{MTJ}}$. We further observe that $V_{\text{MTJ}}$ has a similar reducing impact on $t_0$ in the entire measured range of $V_{\text{SOT}}$. As discussed in Sec.~\ref{5}.A, this reduction can be decomposed into a symmetric and an asymmetric component with respect to $V_{\text{MTJ}}$ = 0, where the latter is ascribed to VCMA. In the diagram of $\Delta t$, we find a less pronounced reduction of its median value toward higher $V_{\text{SOT}}$ and positive $V_{\text{MTJ}}/V_{\text{SOT}}$ ratio. The former is in line with the observation that 1/$\Delta t$ (proportional to the domain-wall velocity in the free layer) increases linearly with the overdrive of $V_{\text{SOT}}$ \cite{Krizakova2020}. The latter can be ascribed to STT, which can assist or hinder the SOT-driven domain-wall motion, and thereby facilitate (impede) the switching at positive (negative) $V_{\text{MTJ}}/V_{\text{SOT}}$ ratio. 

With increasing SOT overdrive, the isocurves (white lines) in the diagrams of $t_0$ become systematically more sparse, since $t_0$ reduces approximately with $1/V_{\text{SOT}}$. Such a scaling characterizes the intrinsic switching regime, which has been observed for $t_{\text{pulse}} < 5$\,ns in our devices \cite{Krizakova2020}. At longer time scales, a transition from the intrinsic to the thermally-activated regime, in which $V_{\text{c}} \propto \sqrt{\ln(t_{\text{pulse}})}$, occurs \cite{Bedau2010,Liu2014a}. We note that, although thermal activation enables switching at $V_{\text{c}}$ lower than the intrinsic critical voltage, this regime is characterized by wide dispersion of the switching onsets [see Fig.~\ref{fig6}(a)], which ultimately limits the reduction of $V_{\text{c}}$. 

These conclusions are supported by the micromagnetic simulations shown in Fig.~\ref{fig8}(b). The simulated diagrams replicate the experimental features with only minor differences. In particular, $t_0$ reduces more abruptly with $V_{\text{MTJ}}/V_{\text{SOT}}$ than in the experiment in those configurations for which VCMA and heating concur. This can be expected because of the fully deterministic nature of our model, which considers neither the stochastic events due to the finite temperature of the environment nor possible dependence of VCMA on temperature. Nevertheless, the scaling with $V_{\text{SOT}}$ accurately portrays the reduction of the characteristic times with the increasing spin current and Joule heating, both generated by $j_{\text{SOT}}$, and the overall trend with $V_{\text{MTJ}}/V_{\text{SOT}}$.


\section{\label{6}Compact model of the effects induced by the MTJ bias}

To quantify the importance of VCMA, STT, and heat on the reduction of $V_{\text{c}}$ in three-terminal devices, we present below a simple phenomenological model of the critical switching conditions, as well as additional measurements of $V_{\text{c}}$ in devices of different sizes that validate this model. The effects induced by the MTJ bias are linked to the geometrical and material parameters of the MTJs, which allows for their separate tuning and eventually provides a starting point to maximize the switching efficiency and reliability of three-terminal devices.

\subsection{Normalized critical voltage for SOT switching}

\begin{figure*}
	\includegraphics[scale=1]{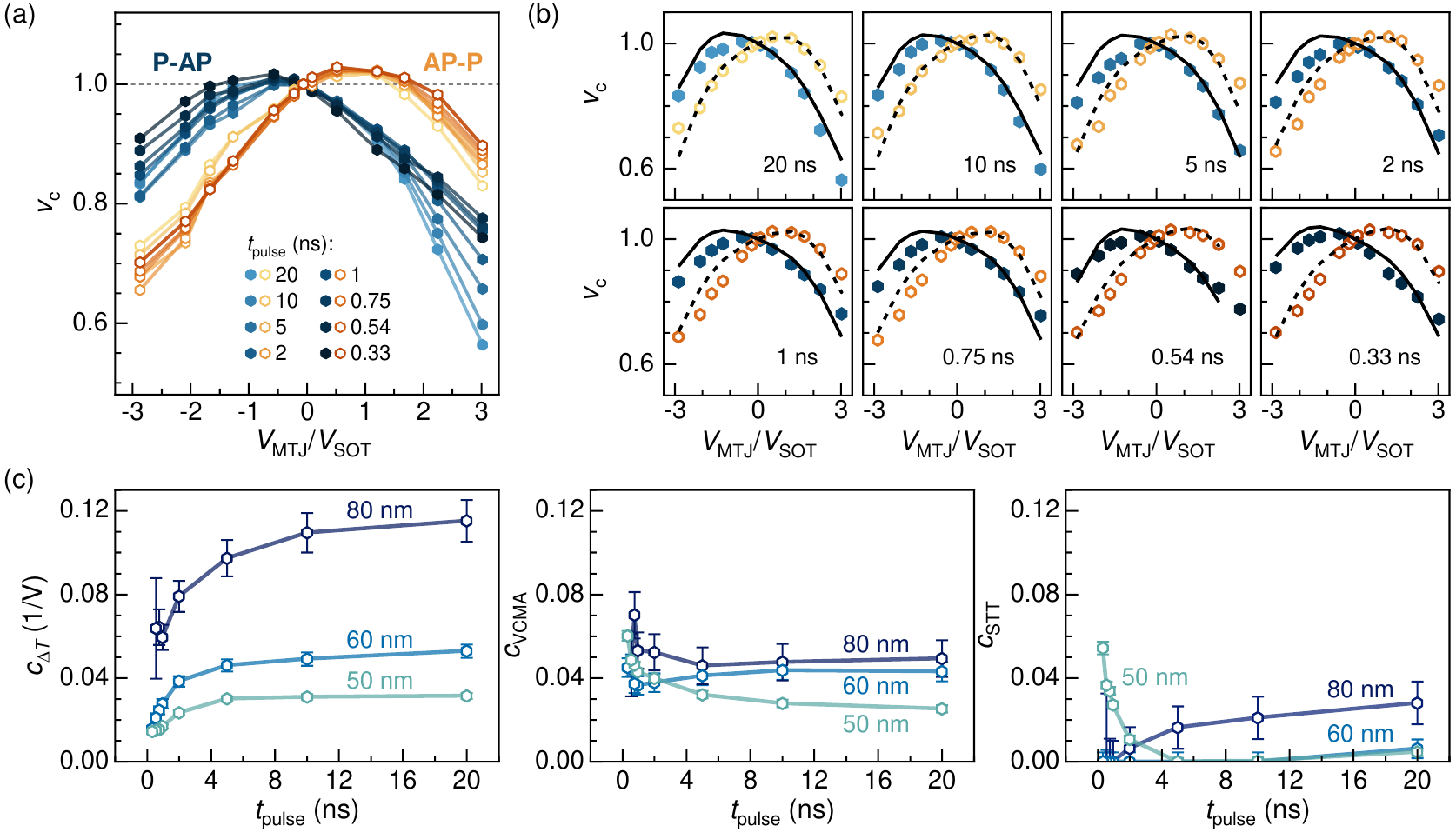}
	\caption{\label{fig9} (a) Normalized critical switching voltage $v_{\text{c}}$ as a function of $V_{\text{MTJ}}/V_{\text{SOT}}$ in the 60\,nm MTJ device. P-AP (AP-P) reversal is shown in a blue (orange) color palette for different $t_{\text{pulse}}$. (b) Fits of $v_{\text{c}}$ obtained for different $t_{\text{pulse}}$ using Eqs.~\ref{eq:3} and \ref{eq:4}. (c) Fitting coefficients $c_{\Delta T}$, $c_{\text{VCMA}}$, and $c_{\text{STT}}$ as a function of $t_{\text{pulse}}$ for three MTJ devices with different diameters. The coefficients represent the relative contribution of Joule heating, VCMA, and STT to the reduction of $v_{\text{c}}$.}
\end{figure*}

As a first step, we separate the effects induced by $V_{\text{SOT}}$ and $V_{\text{MTJ}}$ and account for possible device-to-device variations and configuration-dependent effects, such as $H_{\text{offset}}$. This can be achieved by normalizing the critical switching parameters for each pair of $V_{\text{SOT}}$ and $V_{\text{MTJ}}$ to the critical parameters for switching induced by SOT alone. We define the normalized critical voltage
\begin{equation}\label{eq:1}
v_{\text{c}} = V_{\text{c}}/V_{\text{c0}} = (V_{\text{c0}} + \Delta V_{\text{c}})/V_{\text{c0}},
\end{equation}
where $\Delta V_{\text{c}}$ is the change of $V_{\text{c}}$ attributed to $V_{\text{MTJ}}$ and $V_{\text{c0}}$ is $V_{\text{c}}(V_{\text{MTJ}} = 0)$.  We measured $V_{\text{c}}$ from 200 switching trials measured at each combination of $t_{\text{pulse}}$ and $V_{\text{MTJ}}$. Figure~\ref{fig9}(a) shows $v_{\text{c}}$ as a function of $V_{\text{MTJ}}/V_{\text{SOT}}$ for different pulse widths and both switching configurations. The plot has the same overall symmetry as the real-time measurements of $t_0$ [see Figs.~\ref{fig6}(a) and \ref{fig8}(a)] showing that $v_{\text{c}}$ reduces strongly with increasing $V_{\text{MTJ}}/V_{\text{SOT}}$ due to the concurring effects of heat and VCMA. Depending on the switching polarity, we also observe that intermediate bias conditions result in $v_{\text{c}} > 1$ when the increase of the anisotropy due to VCMA is stronger than its reduction due to heating. Additionally, the plot reveals the dependence of $v_{\text{c}}$ on $t_{\text{pulse}}$, which is most visible for $V_{\text{MTJ}}/V_{\text{SOT}} > 0$. For example, $v_{\text{c}}$ for AP-P switching is reduced by 20\% using 0.3-ns-long pulses and more than 40\% by 20-ns-long pulses. This disparity is assigned to the stronger role played by thermal effects when longer pulses are applied. On the other hand, for $V_{\text{MTJ}}/V_{\text{SOT}} < 0$, the lowest $v_{\text{c}}$ can be achieved at $t_{\text{pulse}} \approx 2$\,ns, which indicates the presence of competing effects in the long-pulse limit. To elaborate further on this point, we devise a simple model that accounts for the dependence of $v_{\text{c}}$ on $V_{\text{MTJ}}$ at different time scales.

\subsection{Compact model of SOT switching assisted by $V_{\text{MTJ}}$}

Based on the results presented in Sec.~\ref{3}-\ref{5}, the effects induced by $V_{\text{MTJ}}$ on $\Delta V_{\text{c}}$ can be divided into three parts related to VCMA, STT, and Joule heating. The VCMA effect, to a first approximation, is described by a linear dependence of the effective magnetic anisotropy ($K$) on $V_{\text{MTJ}}$, with the proportionality factor given by $\varepsilon \propto \text{d}K/\text{d}V_{\text{MTJ}}$, \cite{Nozaki2010,Yoda2017}. Assuming $V_{\text{c}}\propto K$ \cite{Lee2013,Feng2020}, the change of magnetic anisotropy translates to a linear variation of $V_{\text{c}}$ with $V_{\text{MTJ}}$ \cite{Grimaldi2020}. Similarly, the STT induces a linear variation of $V_{\text{c}}$ that is proportional to $V_{\text{MTJ}}$ \cite{Sun2000}. Finally, Joule heating in the system consists of two terms: one proportional to the power dissipated in the MTJ pillar \cite{Mihajlovic2020}, ${V_{\text{MTJ}}}^2/R_{\text{MTJ}}$, and the other to the power dissipated in the SOT track, ${V_{\text{SOT}}}^2/R_{\text{SOT}}$. The latter term, however, influences $V_{\text{c0}}$ and is taken into account by the normalization in Eq.~\ref{eq:1}. We thus assume that $V_{\text{c}}$ reduces quadratically with $V_{\text{MTJ}}$. Following the above reasoning, we express $v_{\text{c}}(V_{\text{MTJ}})$ as

\begin{equation}
	\begin{aligned}\label{eq:2}
v_{\text{c}} ={} & 1 - \left[\frac{2 \varepsilon V_{\text{MTJ}}}{M_{\text{S}} t_{\text{FL}} t_{\text{MgO}}} \pm \frac{\beta V_{\text{MTJ}}}{RA(1 - b|V_{\text{MTJ}}|)} \right. \\
&\left. + \frac{\zeta {V_{\text{MTJ}}}^2}{RA(1 - b|V_{\text{MTJ}}|)}\right] \cdot \left(H_{\text{k}} \mp H_{\text{offset}} \right)^{-1},
	\end{aligned}
\end{equation}

where the upper (lower) signs correspond to the P-AP (AP-P) switching. $H_{\text{k}}$ is the effective anisotropy field, $\varepsilon$ the VCMA coefficient in [J/V\,m], $\beta$ the STT efficiency in [Oe\,m$^2$/A], $\zeta$ the current-induced heating in [Oe\,m$^2$/W], $RA$ is the resistance-area product in the P or AP state, and $b$ is a coefficient used to approximate the bias dependence of the TMR -- equal to 0.4\,(0.7)\,V$^{-1}$ for the P (AP) state in the 60\,nm MTJ. In general, $\beta$ can differ for the P-AP and AP-P reversals and $\varepsilon$ can change depending on the sign of $V_{\text{MTJ}}$ \cite{Wang2012,Stiles2006}. Moreover, the extent to which each effect contributes to $v_{\text{c}}$ is time-dependent. Whereas this is evident for Joule heating, given that the temperature in the MTJ increases with time during a pulse, temporal variations of VCMA and STT can arise as a consequence of heating.

To limit the number of parameters, the model can be simplified by substituting the geometrical and material parameters in Eq.~\ref{eq:2} by $c_{\text{VCMA}}$, $c_{\text{STT}}$, and $c_{\Delta T}$, i.e. coefficients representing the VCMA, STT, and thermal effects, respectively. We allow these coefficients to vary with $t_{\text{pulse}}$, but impose the same value for both switching configurations. With these assumptions, Eq.~\ref{eq:2} for P-AP and AP-P switching reads
\begin{equation}\label{eq:3}
\begin{aligned}
{v_{\text{c}}}^{\text{P-AP}} ={}& 1 - \frac{1}{|V_{c0}^{\text{P-AP}}|} \left( c_{\text{VCMA}} V_{\text{MTJ}} + c_{\text{STT}} V_{\text{MTJ}} \right. \\ &\left. + c_{\Delta T} {V_{\text{MTJ}}}^2  \right),
\end{aligned}
\end{equation}
\begin{equation}\label{eq:4}
\begin{aligned}
{v_{\text{c}}}^{\text{AP-P}} ={}& 1 - \frac{1}{|V_{c0}^{\text{AP-P}}|} \left( c_{\text{VCMA}} V_{\text{MTJ}} - c_{\text{STT}} V_{\text{MTJ}}  \right. \\ &\left. + c_{\Delta T} {V_{\text{MTJ}}}^2  \right).
\end{aligned}
\end{equation} 
Figure~\ref{fig9}(b) shows the results of the fits of $v_{\text{c}}$ as a function of $V_{\text{MTJ}}/V_{\text{SOT}}$ performed using Eqs.~\ref{eq:3} and \ref{eq:4} for different values of $t_{\text{pulse}}$. The agreement between the data and model is very good considering the simplicity of the model; further improvements can be achieved by using different $c_{\text{VCMA}}$ and $c_{\text{STT}}$ coefficients for P-AP and AP-P switching.

\subsection{Size dependence}

Because the amount of dissipated heat scales with device size, we expect a scaling of $c_{\Delta T}$ with the area of the MTJ. To validate our model for different devices we repeated the experiment using two additional MTJs with a diameter of 50\,nm and 80\,nm and a similar $RA$ product as the 60\,nm device presented previously. For each device, we applied a single pulse amplitude, $V_{\text{SOT}} = 1.1V_{\text{c0}}$ for AP-P switching at $t_{\text{pulse}} = 15$\,ns. Figure~\ref{fig9}(c) shows the coefficients $c_{\Delta T}$, $c_{\text{VCMA}}$, and $c_{\text{STT}}$ obtained for the three devices as a function of $t_{\text{pulse}}$. As expected, $c_{\Delta T}$ scales with the MTJ size and increases exponentially with $t_{\text{pulse}}$ from the pulse onset until it saturates at a value that is proportional to the area of the free layer. The saturation is reached faster in the smaller devices, which can be understood by considering that the heat capacity of the MTJ scales with the volume and, once the temperature is constant, a prolongation of $t_{\text{pulse}}$ in the thermally activated regime does not reduce further the critical voltage. 

Also $c_{\text{VCMA}}$ and $c_{\text{STT}}$ increase with the MTJ area, although in a less pronounced way than $c_{\Delta T}$. This variation demonstrates that SOT-induced switching is more affected by $V_{\text{MTJ}}$ when the size of the free layer moves further away from the macrospin limit, that is, when magnetization reversal occurs by domain nucleation and propagation. In the largest MTJ, Joule heating has a larger effect compared to VCMA and STT, as seen by comparing $c_{\Delta T}$, $c_{\text{VCMA}}$, and $c_{\text{STT}}$ in Fig.~\ref{fig9}(c), whereas in smaller devices the three coefficients become comparable. Therefore, the relative importance of VCMA is expected to grow and eventually dominate over heating if the device is further downscaled or the $RA$ increased. STT has a smaller effect than VCMA, even though the $RA$ pro\-duct of the tunnel barrier in our devices ($\approx$20\,$\Omega\,\mu$m$^2$) is by one to two orders of magnitude lower than in typical VCMA-switching devices \cite{Grezes2017,Wu2021,Yoda2017,Kato2018a}. 

Unlike $c_{\Delta T}$, $c_{\text{VCMA}}$ depends weakly on $t_{\text{pulse}}$. The decrease of $c_{\text{VCMA}}$ by about 30\% as $t_{\text{pulse}}$ increases from 0.33 to 4\,ns suggests that the VCMA affects $v_{\text{c}}$ mainly at the time scales of the intrinsic switching regime. This corroborates our assumption that the VCMA lowers the energy barrier for domain nucleation before the magnetic anisotropy is weakened by the temperature increase. In contrast, $c_{\text{STT}}$ is negligible at sub-ns time scales and increases monotonously with $t_{\text{pulse}}$, with the notable exception of the smallest MTJ. Consequently, STT has a negligible impact on $v_{\text{c}}$ in the larger devices unless sufficiently long pulses are applied. These observations suggest that STT does not contribute effectively to domain nucleation, but may promote fast reversal in small devices where the magnetization dynamics is more coherent \cite{Bouquin2018,Meo2021}.

As a final remark, we note that the relative significance of Joule heating, VCMA, and STT on the SOT-induced switching, given by $c_{\Delta T} > c_{\text{VCMA}} > c_{\text{STT}}$, agrees with the dc-limit results, which were obtained in Sec.~\ref{3} from the analysis of the switching fields under the application of $V_{\text{dc}}$. These two methods, therefore, provide complementary information about the effects originating from a voltage bias on the MTJ.


\section{Conclusions}

In summary, we studied the impact of a bias voltage $V_{\text{MTJ}}$ applied to the top electrode of a three-terminal MTJ device during field-free SOT switching at different time scales. We showed that $V_{\text{MTJ}}$ can substantially reduce the critical energy with respect to switching by SOT alone. For $V_{\text{MTJ}}$ smaller than the STT switching threshold, the combination of $V_{\text{MTJ}}$ and $V_{\text{SOT}}$ reaches an energy efficiency comparable to that of STT, without compromising the capability to switch the MTJ by sub-ns pulses, which is a main advantage of SOT.

We identified three effects contributing to the enhancement of the switching efficiency that originate from the simultaneous application of $V_{\text{SOT}}$ and $V_{\text{MTJ}}$ to a three-terminal device, namely current-induced heating, VCMA, and STT. We investigated these effects using real-time detection of the magnetization reversal by ns-long current pulses. Our study indicates that the switching of sub-100 nm MTJs due to the concurrent effects of $V_{\text{SOT}}$ and $V_{\text{MTJ}}$ occurs by domain nucleation assisted by Joule heating and VCMA, and domain-wall propagation driven by SOT, which is weakly assisted or countered by STT. The activation delay, which is the switching rate-limiting factor when the writing current is low, can be efficiently reduced by $V_{\text{MTJ}}$. The conclusions drawn from the experiment are corroborated by micromagnetic simulations that include the effects of time-dependent temperature increase, VCMA, and STT, and determine their separate as well as combined impact on the SOT-induced reversal.

We used post-pulse switching probability measurements in different bias conditions to demonstrate the importance of tuning $t_{\text{pulse}}$ and setting the sign and magnitude of $V_{\text{MTJ}}$ to achieve the most efficient switching. To facilitate the optimization, we introduced a compact model that can be used to separate and predict the effects of $V_{\text{MTJ}}$ on the critical SOT switching voltage. Using MTJs with a diameter of 50, 60, and 80 nm, we found that Joule heating in the MTJ pillar has the strongest influence on the critical voltage in the larger MTJs for $t_{\text{pulse}} > 1$\,ns, whereas the relative importance of VCMA gradually increases in the smaller MTJs, as well as at short time scales. For $V_{\text{MTJ}}$ below the STT switching threshold, STT plays a minor role compared to Joule heating and VCMA, despite the fact that the $RA$ product and the VCMA coefficient are relatively low in our devices. Our findings have general relevance for understanding and optimizing the switching properties of three-terminal MTJ devices, in which multiple current- and voltage-induced effects can be exploited to improve their speed and energetic efficiency.

\begin{acknowledgments}
	
This research was supported by the Swiss National Science Foundation (Grant no. 200020-172775), the Swiss Government Excellence Scholarship (ESKAS-Nr. 2018.0056), the ETH Zurich (Career Seed Grant SEED-14 16-2), and imec’s Industrial Affiliation Program on MRAM devices.
\end{acknowledgments}

\bibliography{Biblio}

\end{document}